\documentclass[12pt]{article}
\usepackage{epsfig}
  
\def\be{\begin{equation}}
\def\ee{\end{equation}}
\def\ben{\begin{displaymath}}
\def\een{\end{displaymath}}
\def\ba{\begin{array}{c}}
\def\bal{\begin{array}{l}}
\def\ea{\end{array}}

\begin{document}

 \begin{center}
{\tiny .}

\vspace{.35cm}

 {\Large \bf
%
%\vspace{.35cm}
Classification of the conditionally observable spectra exhibiting
central symmetry

   }\end{center}

\vspace{10mm}

 \begin{center}

\vspace{5mm}

  {\bf Jun-Hua Chen}

 \vspace{3mm}
%\'{U}stav jadern\'e fyziky AV-\v{C}R,

Department of Physics, FNPE, Czech Technical University,

B\v{r}ehov\'{a} 7, 115 19 Prague, Czech Republic \vspace{3mm}

{e-mail: cjh611@gmail.com
 }

\vspace{6mm}

 {\bf Edita Pelantov\'{a}}

 \vspace{3mm}
%\'{U}stav jadern\'e fyziky AV-\v{C}R,

Department of Mathematics, FNPE, Czech Technical University,

Trojanova 13, 115 19 Prague, Czech Republic \vspace{3mm}

{e-mail: pelantova@km1.fjfi.cvut.cz }

\vspace{3mm}

 and

\vspace{5mm}

 {\bf Miloslav Znojil}

 \vspace{3mm}
%\'{U}stav jadern\'e fyziky AV-\v{C}R,
Nuclear Physics Institute ASCR,

 250 68 \v{R}e\v{z}, Czech Republic

\vspace{3mm}

{e-mail: znojil@ujf.cas.cz}

\vspace{3mm}

\vspace{5mm}
%

%\today, merger.tex

\end{center}

\vspace{5mm}

\newpage

\section*{Abstract}

The emergence of upside-down symmetry of the bound-state energies
$E_1=-E_N$, $E_2=-E_{N-1}$, $\ldots$ has been observed,  in
strong-coupling regime, in several pseudo-Hermitian $N-$state
quantum systems. We show that such a symmetry assumption also
simplifies a combinatorial classification of these systems since
non-equivalent hierarchies of complexifications beyond the
strong-coupling boundaries induce non-equivalent pairwise links
between the energy levels.

\newpage

\section{Introduction}

In 2002,  Dunne and Shifman \cite{BD} noticed and emphasized that
in several quasi-exactly solvable models a number of interesting
mathematical as well as physical consequences can be deduced from
the emergence of the ``central symmetry" duality
 \be
 E_n-c_{(lower)} = c_{(upper)}-E_{N-n}\,,\ \ \ \ n=0, 1, \ldots
 \label{updown}
 \ee
between certain low-lying and  highly-excited bound states. In an
entirely different context, our recent series of papers
\cite{2x2,3x3,maximal,condit} revealed that the same type of the
central symmetry of the spectrum seems to play a decisive role
also during an efficient fine-tuning suppression of instabilities
(or, if you wish, of ``quantum catastrophes") in certain
pseudo-Hermitian phenomenological $N-$site chain models.

In our present short note we intend to broaden the scope of the
latter set of references by paying attention, in principle, to
{\em all} the $N-$dimensional pseudo-Hermitian matrix
Hamiltonians. In this sense we are going to complement some
considerations presented in the latter reference \cite{condit}. In
particular, we shall return to the problem of the classification
of all the possible ``conditionally stable" spectra which exhibit
the ``stability-friendly" symmetry (\ref{updown}). Indeed, only an
incomplete solution of this classification problem has been
offered, by one of us, in ref.~\cite{condit}.

We shall start by recalling some of the basic ideas of some
previous related papers in section \ref{arte}. We emphasize there
that the collapse of pseudo-Hermitian models can be, generically,
mediated by the various alternative mergers and subsequent
complexifications of certain pairs of the energy levels.
Obviously, a necessity emerges of a correct numbering of the
non-equivalent ``pairings" and possible complexifications.

Our present main result presented in section \ref{linkage} offers
the answer. Some of its consequences will be summarized in section
\ref{summary} where we re-emphasize that the introduction of the
concept of the conditional observability in \cite{condit} opened a
rich variety of new scenarios of collapse of phenomenological
quantum systems.

\section{Confluences of the levels
 in  $2J-$state models  \label{arte} }

\begin{figure}[t]                     %instead of \begin{figure}[t]
\begin{center}                         %instead of \begin{center}
\epsfig{file=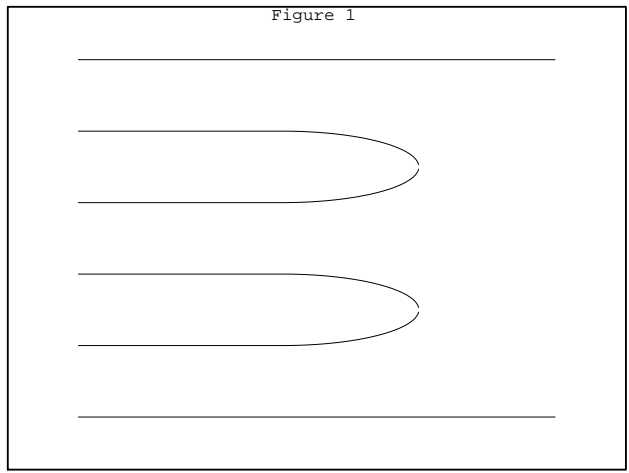,angle=270,width=0.6\textwidth}
\end{center}                         %instead of \end{center}
\vspace{-2mm} \caption{A sample of the mutual attraction of some
of  the energies followed by their confluence and
complexification.
 \label{obr1a}}
\end{figure}

In a typical ``catastrophic" scenario as sampled in
ref.~\cite{condit} the growth of a parameter in the Hamiltonian
$H$ forces some neighboring energy levels to merge and,
subsequently, to complexify. A schematic picture of such a
situation is provided here by Figure \ref{obr1a} where six levels
of some energy spectrum $\{E_n\}_{n=1, 2, \ldots,6}$ are displayed
and where the external levels $E_1$ and $E_6$ are assumed
decoupled. We see that when the strength of the non-Hermiticity
reaches a critical ``exceptional point" value \cite{Kato}, the
quantity $E_2$ merges with $E_3$ while $E_4$ merges with $E_5$.

\begin{figure}[t]                     %instead of \begin{figure}[t]
\begin{center}                         %instead of \begin{center}
\epsfig{file=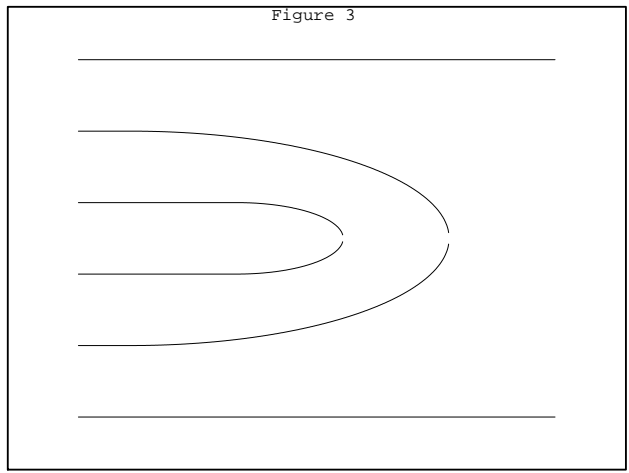,angle=270,width=0.6\textwidth}
\end{center}                         %instead of \end{center}
\vspace{-2mm} \caption{An alternative to Figure \ref{obr1a}.
 \label{obr2b}}
\end{figure}

Another type of a  ``catastrophic" scenario can be produced when
the attraction between $E_3$ and $E_4$ dominates. The resulting
new forms of the parametric dependence of the spectrum have been
sampled in ref.~\cite{condit} (cf. Figure Nr. 5 there). Here, the
message is repeated by Figure \ref{obr2b} which shows that the
levels $E_3$ and $E_4$ merge more quickly than the remaining pair
of the levels $E_2$ and $E_5$.

\begin{figure}[t]                     %instead of \begin{figure}[t]
\begin{center}                         %instead of \begin{center}
\epsfig{file=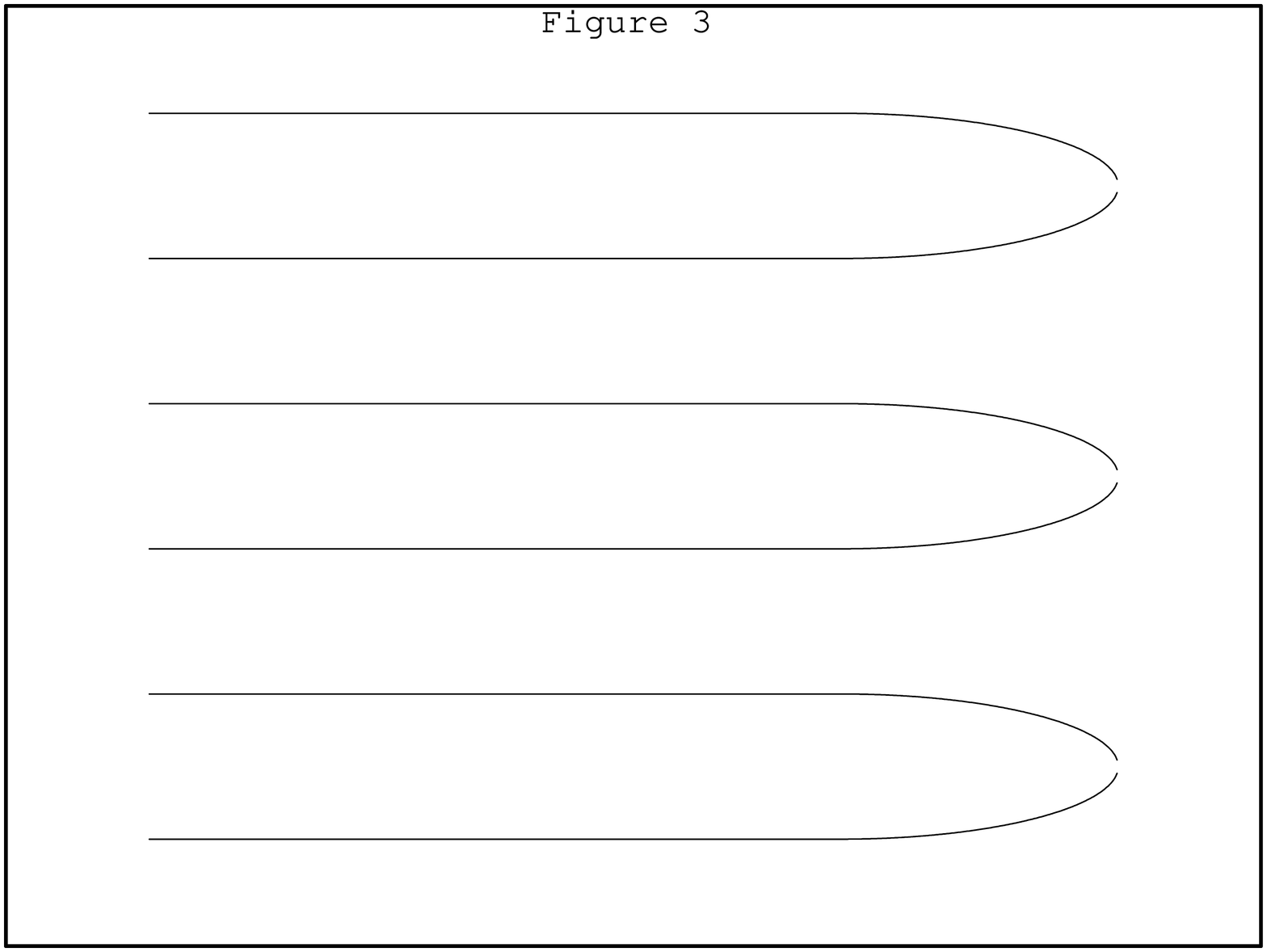,angle=270,width=0.6\textwidth}
\end{center}                         %instead of \end{center}
\vspace{-2mm} \caption{Another alternative to Figure \ref{obr1a}.
 \label{obr3b}}
\end{figure}

In Figure \ref{obr3b} we illustrate the most current scenario
where the spectrum splits in the separate doublets of levels, each
of which reaches its point of confluence separately. Such a
scenario has been found to occur in many toy examples. For
illustration one could recollect, e.g., the exactly solvable
parametric-dependence of the spectrum of the so called ${\cal
PT}-$symmetric harmonic oscillator \cite{ptho} and/or of some of
its simplest differential-equation generalizations \cite{Trinh}.
In this context, the merging levels are sometimes characterized by
the opposite ``quasiparities"  \cite{ptho,geza}.

We can formulate our first conclusion that while there is just one
possible confluence pattern in two-state models \cite{2x2}, the
generic four-state models already admit the two distinct
complexification patterns as sampled in Figures \ref{obr1a} and
\ref{obr2b}. Similarly, in all the ``next", six-state models one
has to contemplate the three substantially different forms of the
instability of the system. Their respective complexification
patterns can easily be read out of our Figures \ref{obr1a} -
\ref{obr3b}. In the first two cases it is only necessary to
imagine that the effect of the ``hidden" attraction only affects
the external levels $E_1$ and $E_6$ somewhere very far to the
right, i.e., out of the range of our pictures.

At the higher even dimensions $N=2J$, the level-linking patterns
become perceivably more complicated. Still, up to the dimension as
large as $N=2J=14$, one of us showed in ref.~\cite{condit} that
their classification may still be based on a simplified recurrence
relation. In what follows we intend to extend the latter analysis
to and beyond the dimension $N=16$. We shall see that the
exhaustive and complete enumeration of {\em all} the eligible
arrangements of the quantum catastrophes remains feasible and that
it may be obtained in closed form, provided only that a
generating-function approach to the problem is employed.

\section{Counting the non-equivalent scenarios  \label{linkage} }

 \subsection{Notation}

Modifying slightly the notation conventions of ref.~\cite{condit}
let us consider a generic spectrum $E_1 < E_2 < \ldots < E_{2J}$.
Let us further assume that sooner or later all these energies
complexify, pairwise, with the growth of a suitable parameter
$\lambda$. Our task will be an explicit description of the
eligible possibilities. Once the symbol $P^{(2J)}$ denotes the
number of the non-equivalent confluence pairings of the levels, we
may recollect that $P^{(2)}=1$, $P^{(4)}=2$ (cf. Figures
\ref{obr1a} and \ref{obr2b} once more) and $P^{(6)}=3$.

All the multiple mergers will be ignored as mere trivial limiting
cases. Thus, in the slightly modified shorthand notation of
ref.~\cite{condit} the unique merger pattern  at $J=1$ will be
characterized by the compressed level-number symbol $\{[1,2]\}$.
Similarly, at $J=2$ the two symbols $\{[2,3],[4,5]\}$ and
$\{[2,5],[3,4]\}$  will be assigned to the two respective
possibilities (cf. the inner parts of Figures \ref{obr1a} and
\ref{obr2b}). Finally, the models with $J=3$ will be assigned the
symbols $\{[1,6],[2,3],[4,5]\}$, $\{[1,6],[2,5],[3,4]\}$  and
$\{[1,2],[3,4],[5,6]\}$, etc.

\begin{table}[b]
\caption{Multiplicities $P^{(2J)}$ of the merging patterns}
\label{pexp2}
\begin{center}
\begin{tabular}{||c|cccccccccc||}
\hline \hline
 $J$&0&1&2&3&4&5&6&7&8&
 %10&
 %11&
 \ldots\\
 \hline
 $P^{(2\,J)}$&
 1&1&2&3&6&10&20&35&70&\ldots\\
 \hline \hline
\end{tabular}
\end{center}
\end{table}

\subsection{Classification}

%$P^{(2J)}=C_{J}^{[\frac{J}{2}]}$

The set of the lower estimates of the quantities $P^{(2J)}$ as
given in Table Nr. 1 of ref.~\cite{condit} can be replaced by the
following complete and exact result.

\vspace{.15cm}

 {\it \bf Theorem.}
%\vspace{.35cm}

 \noindent
At any integer  $J = 1, 2, \ldots$, the number of the
non-equivalent patterns of the pairwise mergers (i.e., confluences
and subsequent complexifications) of the energy levels of our
$2J-$state model is given by closed formula,
 $$P^{(2J)} = \left (
  \ba
  J \\
  \left [ \frac{J}{2}
  \right ]
  \ea
  \right )\,.
 $$

\vspace{-.15cm}

 {\it \bf Proof.}

%\vspace{.35cm}
 \noindent
First let us lift the requirement of symmetry, and denote the
number of nonequivalent connections as $T^{(2J)}$. As long as the
first energy level can merge with any level $2i$, their link
divides the rest of $2J-2$ levels into two disconnected groups of
$2i-2$ and $2J-2i$ levels. We can write
%\begin{eqnarray}
 \ben
T^{(2J)}=\sum_{i=1}^{J}T^{(2i-2)}T^{(2J-2i)}
=\sum_{i=0}^{J-1}T^{(2i)}T^{(2j-2i-2)}\,.\label{e1}
 %\end{eqnarray}
 \een
For the generating function %$f(x)$ of $T^{(2J)}$ as
\begin{equation}
f(x)=\sum_{J=0}^{\infty}T^{(2J)}x^J
\end{equation}
we have
 \ben
f^2(x)=\sum_{n=0}^{\infty}\sum_{i=0}^{n}T^{(2i)}T^{(2n-2i)}x^n
=\sum_{n=0}^{\infty}T^{(2n+2)}x^n=\frac{f(x)-1}{x}\,
 \een
so that we may conclude that
 \ben
f(x)=\frac{1-\sqrt{1-4x}}{2x}\,,\ \ \ \ \ \ \
T^{(2J)}=\frac{(2J)!}{(J+1)!J!}\,.
 \een
In the next step let us calculate $P^{(2J)}$. Firstly, we can
write down the recurrences
 \ben
P^{(2J)}=P^{(2J-2)}+\sum_{i=1}^{[\frac{J}{2}]}T^{(2i-2)}P^{(2J-4i)}
%\nonumber\\
=P^{(2J-2)}+\sum_{i=0}^{[\frac{J}{2}]-1}T^{(2i)}P^{(2J-4i-4)}
 \een
which add the ``missing terms" in the simplified lower-estimate
relations as employed in ref.~\cite{condit}. Subsequently, we may
employ again the generating function $g(x)$ of $P^{(2J)}$ and
deduce that
\begin{equation}
g(x)=\sum_{J=0}^{\infty}P^{(2J)}x^J=
1+\sum_{J=1}^{\infty}P^{(2J)}x^J
\end{equation}
where
 \ben
\sum_{J=1}^{\infty}P^{(2J)}x^J=\sum_{J=1}^{\infty}P^{(2J-2)}x^J+
\sum_{J=2}^{\infty}
\sum_{i=0}^{[\frac{J}{2}]-1}T^{(2i)}P^{(2J-4i-4)}x^J\,.
 \een
This enables us to conclude that
 \ben
g(x)-1=x\,g(x)+x^2f(x^2)g(x)\,.
 \een
After all the insertions we arrive at the final formula
 \ben
g(x)=\frac{1}{1-x-x^2f(x^2)}=
\frac{1}{\sqrt{1-4x^2}}+\frac{1-\sqrt{1-4x^2}}{2x\sqrt{1-4x^2}}\,.
 \een
Our final result $P^{(2J)}=C_{J}^{[\frac{J}{2}]}$ (sampled in
Table 1) immediately follows. {\it \bf QED}.

\section{Summary \label{summary} }
%
%In this context, the problem of the correct numbering of the
%non-equivalent ``pairings" and possible complexifications . Our
%main motivation is that only an incomplete solution of this
%classification problem has been offered recently, by one of us, in
%ref.~\cite{condit}.

In our present short note we intended to show that and how the
purely combinatorial classification of all the possible
confluences of the energy levels (i.e., of all the possible
quantum catastrophes) can be performed in full generality. We did
not work with any particular Hamiltonians this time, noticing only
that for constructive purposes, the chain models of
refs.~\cite{maximal} exhibiting central symmetry (\ref{updown})
would prove particularly suitable and friendly in technical sense
again.

A remark can be added that after the extremely popular choice of
the differential-operator Hamiltonians one often has to refrain
just to the most elementary and exceptional
``neighboring-level-confluence" scenario \cite{Trinh}. On the
basis of our present abstract classification scheme one should
expect that the emergence of many less trivial scenarios of the
collapse cannot be excluded even in the differential-operator toy
models. By sophisticated numerical techniques, some explicit
examples have already been discovered recently \cite{DDTb,Aryeh}.

%
%There is always a conflict between the consequent ``first
%principles" of the theory and the purely pragmatic requirements of
%a good fit $\{E_n^{(experimental)}\}\approx
%\{E_n^{(theoretical)}\}$, say, at and $N-$plet of indices $n=1, 2,
%\ldots, N$. For this reason, new classes of models are still being
%developed. Among them, our attention has been paid here to the so
%called quasi-Hermitian Hamiltonians, the Hermiticity of which
%(required by the principles of Quantum Mechanics) is {\em defined}
%in an unusual manner  \cite{Geyer}.
%
%Our present message simply says that Table Nr. 1 of ref.
%\cite{condit} has to be replaced by the present Table \ref{pexp2}.

\vspace{5mm}

\section*{Acknowledgement}

Work supported by the M\v{S}MT ``Doppler Institute" project Nr.
LC06002.

\vspace{5mm}

\section*{Table captions}

\subsection*{Table  \ref{pexp2}.
Multiplicities $P^{(2J)}$ of the merging patterns}

\vspace{5mm}

\section*{Figure captions}

\subsection*{Figure  \ref{obr1a}.
A sample of the mutual attraction of some of the energies followed
by their confluence and complexification.
 }

\subsection*{Figure  \ref{obr2b}. An alternative to Figure \ref{obr1a}.
 }

\subsection*{Figure  \ref{obr3b}. Another alternative to Figure \ref{obr1a}.
 }

\end{document}